\documentclass[twocolumn,prb,showpacs,superscriptaddress,showkeys,preprintnumbers,amsmath,amssymb]{revtex4}

\usepackage{graphicx}
\usepackage{dcolumn}
\usepackage{bm}
\usepackage{multirow}
\usepackage{subscript}
\usepackage{nicefrac}
\usepackage{braket}
\usepackage{amsmath}
\usepackage{verbatim}
\DeclareMathOperator{\Tr}{Tr}

\begin{document}


\title{\textsc{Mag2Pol}: A program for the analysis of spherical neutron polarimetry, flipping ratio and integrated intensity data  }

\author{N. Qureshi}

\email[Corresponding author. Electronic
address:~]{qureshi@ill.fr} 
\affiliation{Institut Laue-Langevin, 71 avenue des Martyrs, CS 20156,  38042 Grenoble Cedex 9, France}

\date{\today}

\begin{abstract}

\textsc{Mag2Pol} is a graphical user interface program which is devoted to the analysis of data from polarized neutron diffractometers with spherical polarization analysis. Nuclear and magnetic structure models can be introduced using space group symbols and individual symmetry operators, respectively, and viewed in an \textsc{OpenGL} widget. The program calculates nuclear/magnetic structure factors, flipping ratios and polarization matrices for magnetic Bragg reflections taking into account structural twins and magnetic domains. Spherical neutron polarimetry data can be analyzed by refining a magnetic structure model including magnetic domain populations in a least-squares fit and can also be correlated to an integrated intensity data set in a joint refinement. Further features are the simultaneous refinement of nuclear and magnetic structures with integrated intensity data and the analysis of flipping ratios either with tabulated magnetic form factors or using a multipole expansion of the magnetization density.

\end{abstract}

\maketitle

\section{Introduction}
\label{sec:Introduction}

The scattering of polarized neutrons from a single-crystalline antiferromagnetic sample with the subsequent full three-dimensional reconstruction of the final neutron spin is known as spherical neutron polarimetry (SNP) [see Refs.~\onlinecite{bro2001,chatt5} for a detailed review] and is a powerful tool for the investigation of complex magnetic structures \cite{bro1991}, chiral magnetic domain populations \cite{bro2005,bab2012} and absolute handednesses. The instrumental realization of such a task has been achieved by the Cryogenic Polarization Analysis Device (CRYOPAD) \cite{tas1989,tas1999,lel2005} which provides an absolutely magnetic field free region at the sample position and therefore full control of the neutron spin by adjusting the magnetic fields strengths and directions at the entrance and exit of the device, which are separated from the sample chamber by an inner Meissner shield provided by a superconducting material.\newline 
A SNP experiment consists in analyzing the components of the final neutron polarization in dependence on the initial one, which yields the so called polarization matrix. The latter is rather a pseudo matrix or a tensor, because it consists of a rotational part to which a vector of created or annihilated polarization is added. The cross section of the neutron with the sample and the change of the incident neutron polarization at the scattering process are expressed by the Blume-Maleev equations \cite{blu1963,mal1963}, which allow to extract important information concerning the magnetic moment configuration. However, the application of this formalism is rather daunting for an unexperienced neutron scatterer and more importantly up to date there exists no computer program based on a graphical user interface, which would allow to analyze SNP data. Several command-line tools like e.g. \textsc{SNPSQ} from the Cambridge Crystallography Subroutine Library (CCSL) \cite{ccsl}, \textsc{MuFit} \cite{poo2012} or modules from the Crystallographic Fortran 95 Modules Library (CrysFML) \cite{crysfml} in principle allow the treatment of SNP data, while \textsc{SORGAM} (again from the CCSL) and \textsc{FullProf} \cite{fullprof} are renowned programs for the analysis of flipping ratio data.  \textsc{Mag2Pol} offers an intuitive interaction with the user for setting up magnetic structure models, which then can be refined to polarization matrices of magnetic Bragg reflections in a least-squares fit. Since the SNP technique is very sensitive to the magnetic moment direction, but not to its magnitude (if no nuclear contribution is present) \textsc{Mag2Pol} provides the unique possibility to combine SNP data with integrated intensity data sets from single-crystal diffractometers in a joint refinement. The weighted agreement factor can be adapted by the user to emphasize a given data set over another. \newline Configurational, orientational and chiral magnetic domains (see Ref.~\onlinecite{chatt5} for a detailed explanation) can easily be introduced and viewed, while their corresponding populations can be included in the refinement process. Structural twins can be included and their populations can be refined to structural data (together with the usual structural parameters like atomic positions, temperature factors, occupations, extinction), which then allows one to use a magnetic structure model with up to eight structural twins combined with up to eight magnetic domains resulting in a maximum of 64 magnetic configurations.\newline
\textsc{Mag2Pol} is also able to treat data obtained by the flipping ratio method, in which the sample is magnetized along the axis of incident neutron polarization and the ratios of diffracted intensities for incident spin-up and spin-down neutrons is measured yielding very precise magnetic structure factors in the case where the nuclear structure is known. Hereby, spherical magnetization densities may be assumed by using tabulated magnetic form factors or the deviation from spherical symmetry can be achieved by a multipole expansion of the magnetization density \cite{bro1979}.\newline \textsc{Mag2Pol} supports of course the refinement of nuclear and magnetic structures on integrated intensity data alone, therfore, covering the major part of neutron single-crystal diffraction methods.  

\section{Graphical user interface and workflow}

\begin{figure}
	\centering
	\caption{\textsc{Mag2Pol}'s graphical user interface after setting up a magnetic structure model.}
	\label{fig:gui}
	\includegraphics[width=0.48\textwidth]{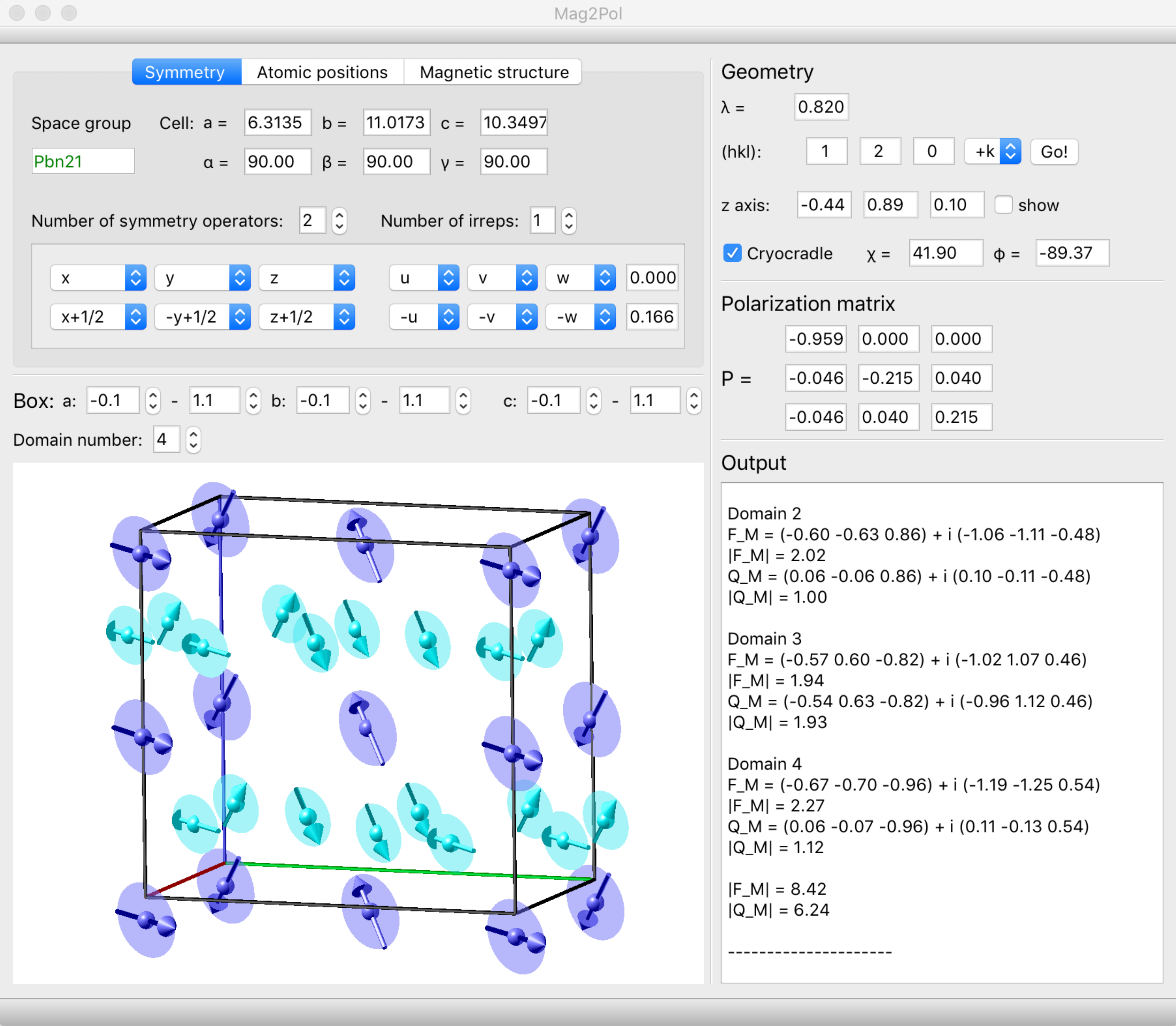}
\end{figure}

The graphical user interface of \textsc{Mag2Pol}'s main window (see Fig.\ref{fig:gui}) is divided into four parts. In the upper left a tab widget contains all the necessary information for the nuclear and magnetic structure models. Those are divided into \textit{Symmetry}, \textit{Atomic positions} and \textit{Magnetic structure}. The structure models are rendered in the \textsc{OpenGL} widget in the lower left part of the window, where the bounding box of the structure and the number of magnetic domain to plot can be chosen. The upper right corner contains the definition of the instrument geometry, where also calculations for single Bragg reflections are triggered. The results of the calculation are shown below in the \textit{Polarization matrix} section and in the \textit{Output} box. The latter will show the nuclear structure factor and the magnetic structure factor for each twin and magnetic domain, respectively. In case of a zero propagation vector the flipping ratio is shown as well (see Sec.~\ref{sec:frs} for mathematical information). \newline

\subsection{Setting up a structure model}

\begin{figure}
	\centering
	\includegraphics[width=0.23\textwidth]{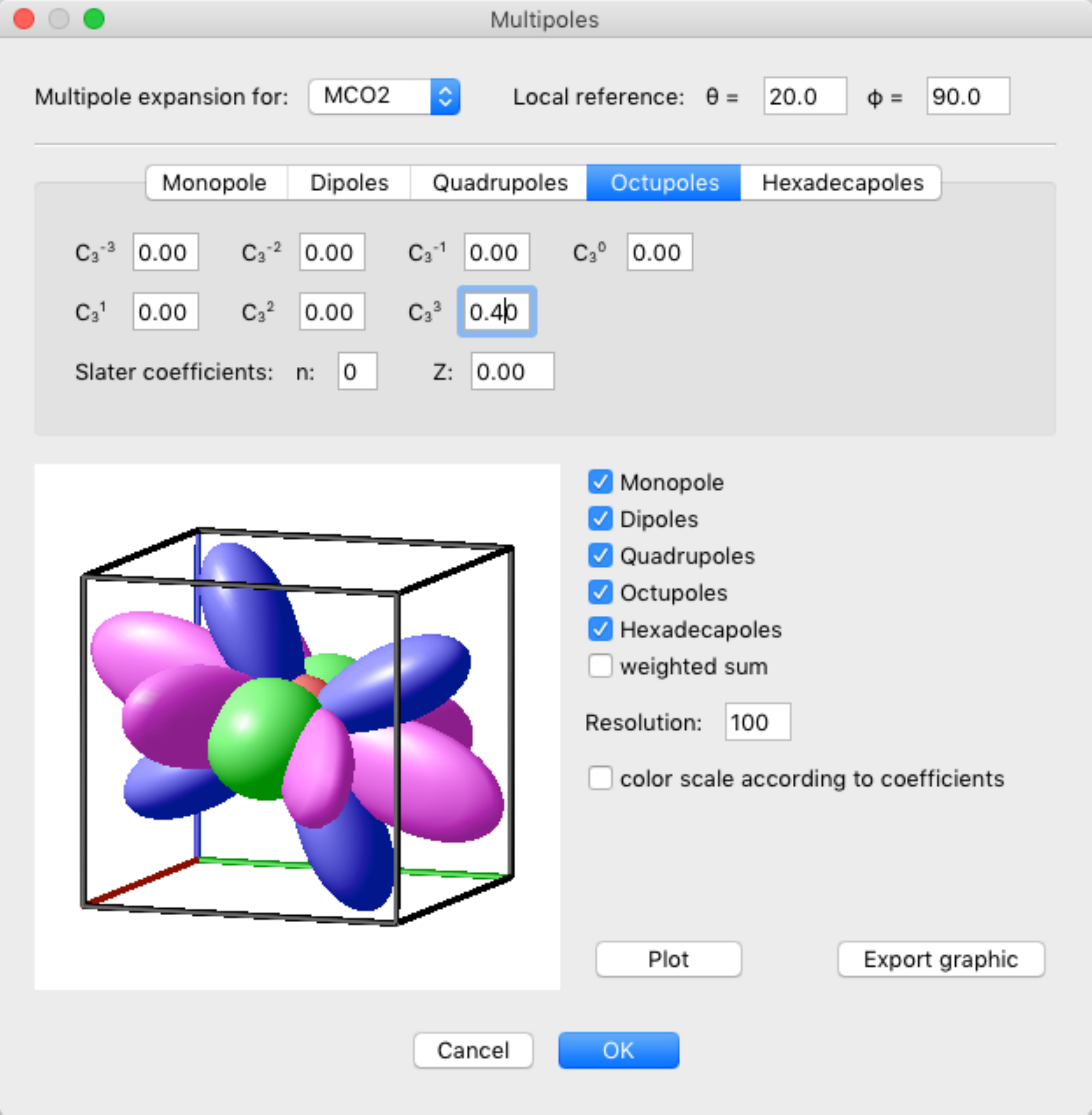}
	\includegraphics[width=0.23\textwidth]{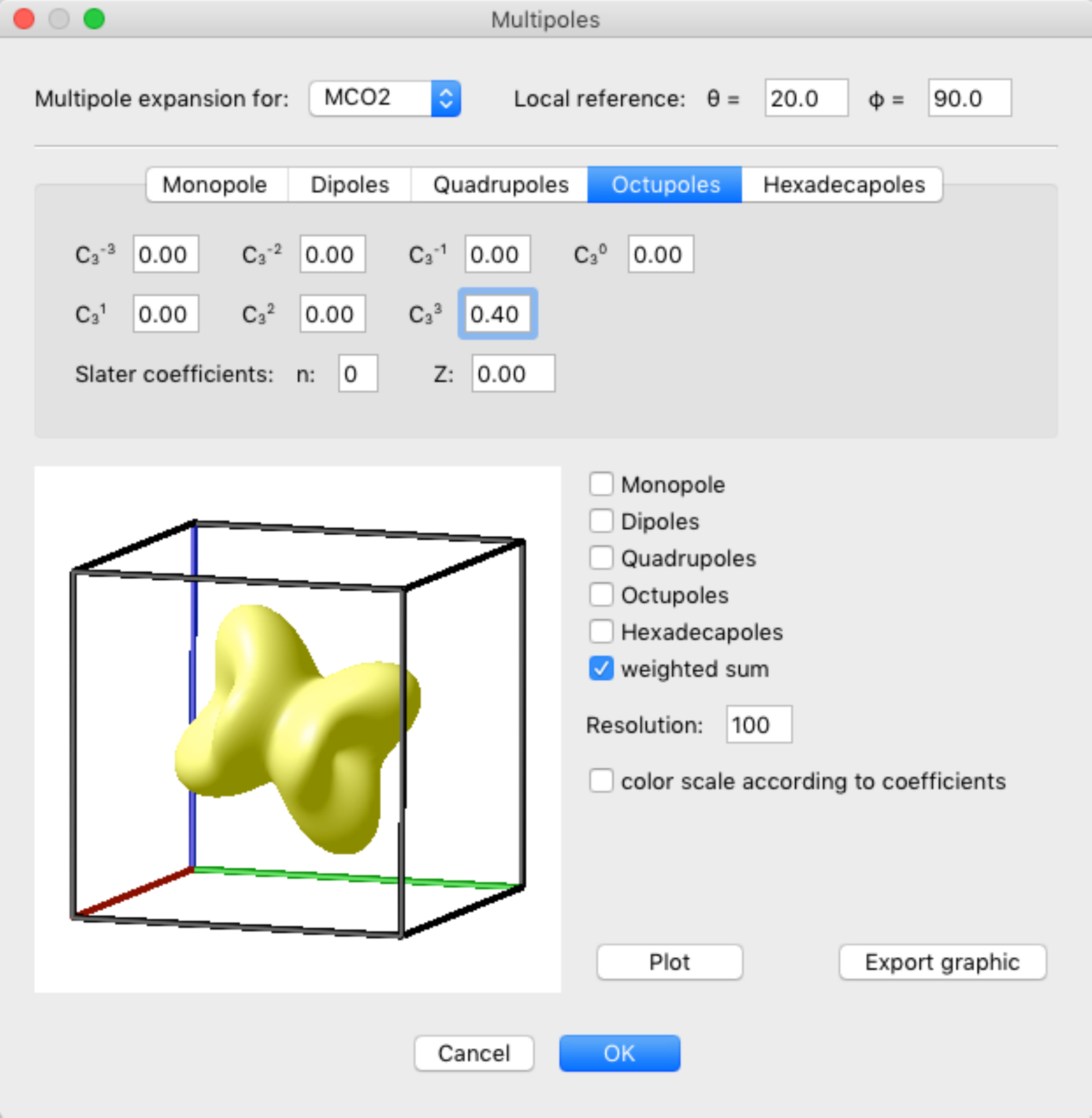}
	\includegraphics[width=0.48\textwidth]{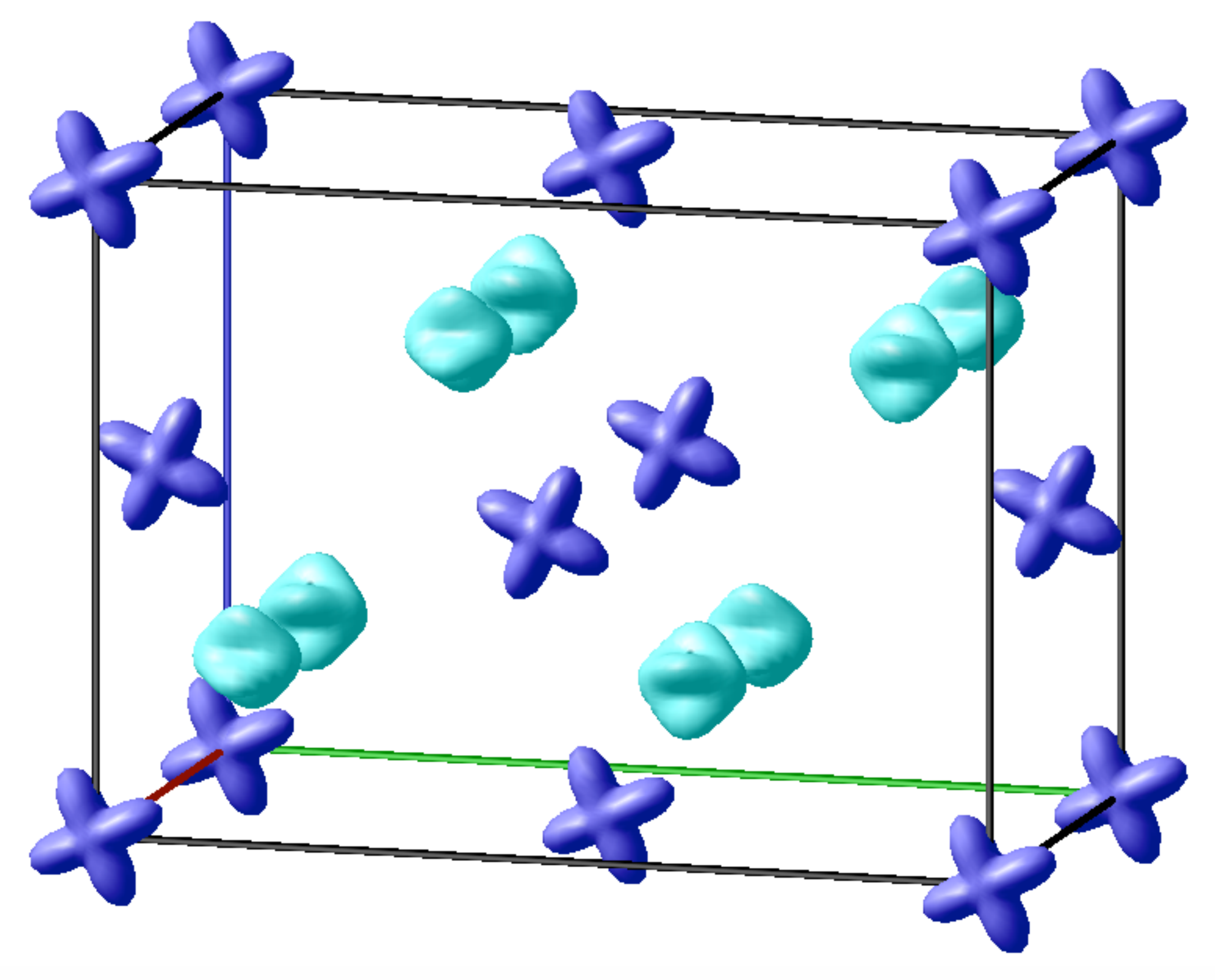}
	\caption{Upper row: Multipole window allowing to the set the coefficients of the multipole expansion of the magnetization density. Upper left: Visualization of the individual orbitals. Upper right: Weighted sum of the individual orbitals. Bottom: Multipole expansion of the magnetization density for each magnetic ion in the crystallographic unit cell with symmetrical constraints given by the space group.}
	\label{fig:mpol_window} 
\end{figure}

The first step for a meaningful use of \textsc{Mag2Pol} is the parametrization of a magnetic structure model. The space group of the underlying nuclear structure has to be entered in Hermann-Mauguin notation (the generators of the 230 space groups in standard setting are stored as well as a few with an additional alternative settings) in the corresponding field under the \textit{Symmetry} tab. The entry is independent of case and additional spaces, and will be shown in green, if the space group has been recognized, and in red in the contrary case. The lattice parameters should be given in \AA ngstr\"oms and degrees. The magnetic symmetry should be given by a number of symmetry operators which can be constructed from the drop-down menues. The first part featuring $xyz$ refers to the symmetry operator which is applied to the atomic position, whereas the second part featuring $uvw$ is the corresponding symmetry operation acting on the magnetic moment. Each symmetry operator can be combined with a magnetic phase. Correct magnetic symmetries based on representation theory can be obtained e.g. from \textsc{BasIreps} in the \textsc{FullProf} suite \cite{fullprof}.\newline
Under the tab \textit{Atomic positions} up to 40 sites can be entered by adjusting the corresponding number. Each site needs an atom label which corresponds to the element symbol in case of a purely nuclear scatterer. The scattering lengths are taken from Ref.~\onlinecite{sea1992}. For a magnetic atom, the label starts with an M followed by the element symbol and the oxidation state, e.g. MCo3. The M refers to the magnetic form factor containing only spherical Bessel functions $j_0$ and is calculated according to an analytical approximation \cite{lis1971,bro2004}. The coefficients for the analytical approximation as well as the scattering length can be accessed from the \textit{Form factors} menu which allows also the plotting and comparison of different magnetic form factors. \textsc{Mag2Pol} allows the creation of user-defined atoms by entering the scattering length, the absorption cross section and the coefficients for the analytical approximation of the magnetic form factor. Alternatively - and this is very useful for crystallographic sites shared by different magnetic ions - a linear combination of an unlimited number of magnetic form factors can be calculated. If the data allows for a more sophisticated description of the magnetic form factor, the deviation of the magnetization density from the spherical symmetry can be achieved by a linear combination of basis functions. In \textsc{Mag2Pol} those basis functions are the real spherical harmonics which describe the angular dependence of the magnetization density. The radial dependence is calculated according to a Slater-type function (see Sec.~\ref{sec:mpoles} for a complete description). The description of each combination of multipoles is always in relation to an orthogonal local reference system, where for $\theta = \varphi = \chi = 0$ the local $x$ axis is parallel to the unit cell's $a$ axis, $y$ is within the $a$-$b$ plane and $z$ is perpendicular to $a$ and $b$. On opening the multipoles window via \textit{Form factors$\rightarrow$Multipoles} the coefficients of the different basis functions (multipoles) can be set together with the Slater-type function coefficients and the local reference angles for each magnetic ion. Non-magnetic ions, e.g. O$^{2-}$ can in principle be treated by first creating a new magnetic ion as described previously and then using that ion in the multipole expansion. \newline
As soon as a coefficient $C_l^m$ is different from zero the respective multipole can be plotted in the \textsc{OpenGL} widget (see Fig.~\ref{fig:mpol_window}, upper left panel). The multipoles are either colored according to their type or to their respective coefficient. It is also possible to visualize the weighted sum of the individual multipoles, which gives an idea of the magnetization density distribution (see Fig.~\ref{fig:mpol_window}, upper right panel). The weighted sum of the multipole population can be visualized in the unit cell instead of spheres (see Fig.~\ref{fig:mpol_window}, bottom) by choosing the corresponding objects to plot in the settings menu. A table containing the so called picking rules according to Ref.~\onlinecite{kar1981} can be invoked from within the program in order to assist the user in the selection of those $C_l^m$ coefficients respecting the site-symmetrical constraints.  \newline 	

Similar to the space group symbol, the entry of the atom label is case independent, however, no spaces can be used. An isotropic temperature factor $B$ and an occupation factor (1 for a fully occupied site) can be entered for each site. Anisotropic displacement parameters can be introduced by right-clicking the isotropic temperature factor of a given atom opening a context menu from which the type (isotropic/anisotropic) can be defined and the anisotropic temperature factors can be set. The displacement ellipsoids can be visualized individually or within the unit cell.\newline
If a magnetic ion has been recognized, it will be displayed in the \textit{Magnetic structure} tab, where its complex Fourier components and a phase factor can be entered. Alternatively, basis vectors can be defined using the corresponding \textit{Structure} menu entry, in which case the parameters in the \textit{Magnetic structure} tab are the coefficients with which the up to six basis vectors are multiplied. In addition, the propagation vector can be entered here, which relates the translational symmetry of the magnetic cell to the nuclear one. The checkbox \textit{in BZ} should be activated, when a non-zero propagation vector is at the interior of the first Brillouin zone. \newline
Once all necessary parameters have been entered, the nuclear and magnetic structures can be rendered in the interactive \textit{OpenGL} widget. In case of a missing or erroneous input, the corresponding error message will pop up. The bounding box of the volume to plot can be modified in the spin boxes corresponding to relative units of the lattice constants. Each individual domain can be viewed by selecting the respective domain number in the spin box. The appearance of every atom/spin site can be modified with the plot options shown at the right of the atomic position. \newline
All parameters which can be set in \textsc{Mag2Pol} can be saved into an input control file in \textsc{ASCII} format (\texttt{*.dat} extension) and can be reloaded at any time. The program also supports loading \texttt{*.cif} and \texttt{*.mcif} files.

\subsection{Calculations}
\begin{figure}
	\centering
	\caption{Upper panel: Intensity map in the $(hk0)$ plane. Lower panel: Projection onto the $(h20)$ line of the map shown in the upper panel by integrating an intensity strip $\pm\Delta k$ around the reflections.}
	\includegraphics[width=0.48\textwidth]{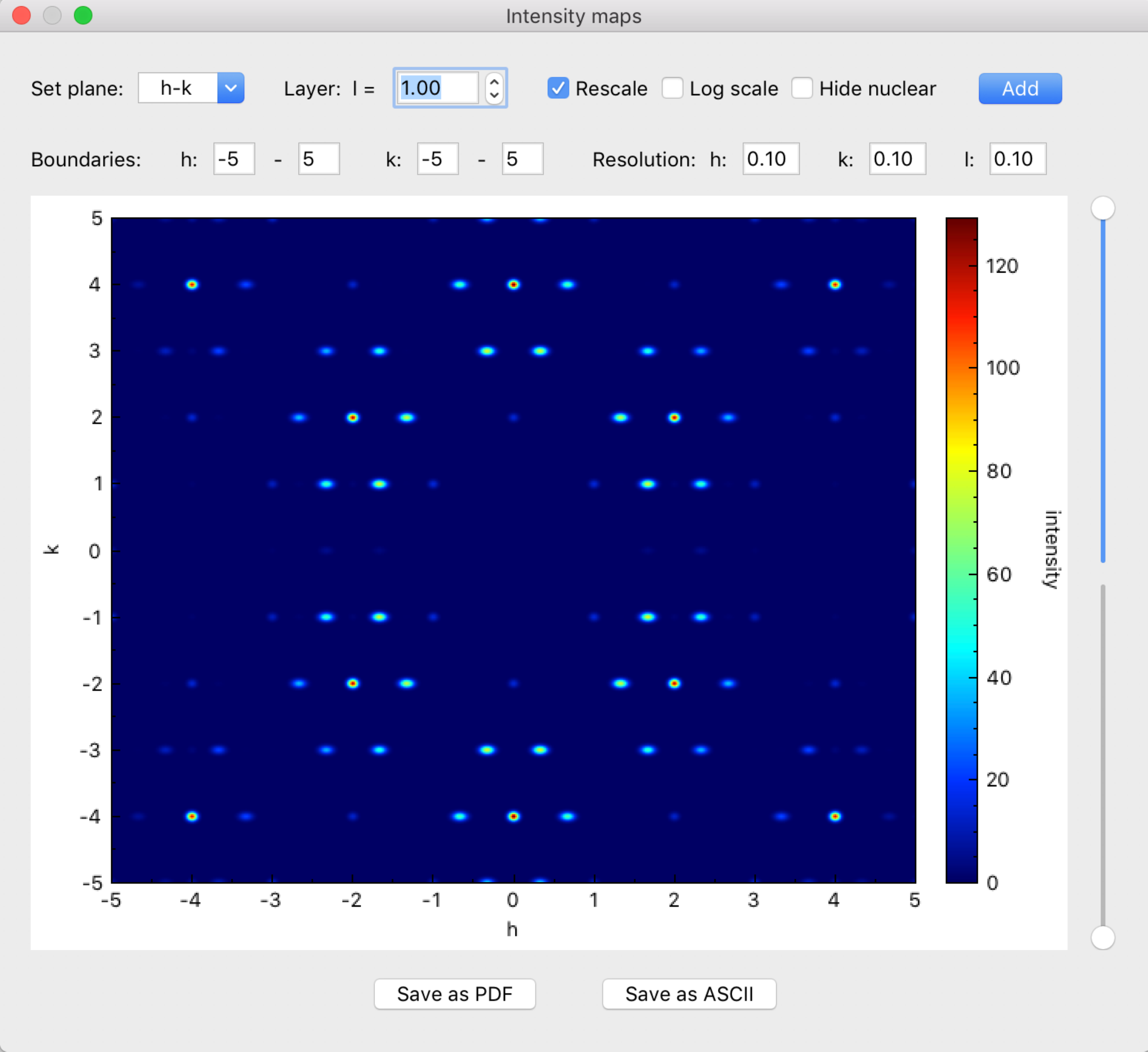}
	\includegraphics[width=0.48\textwidth]{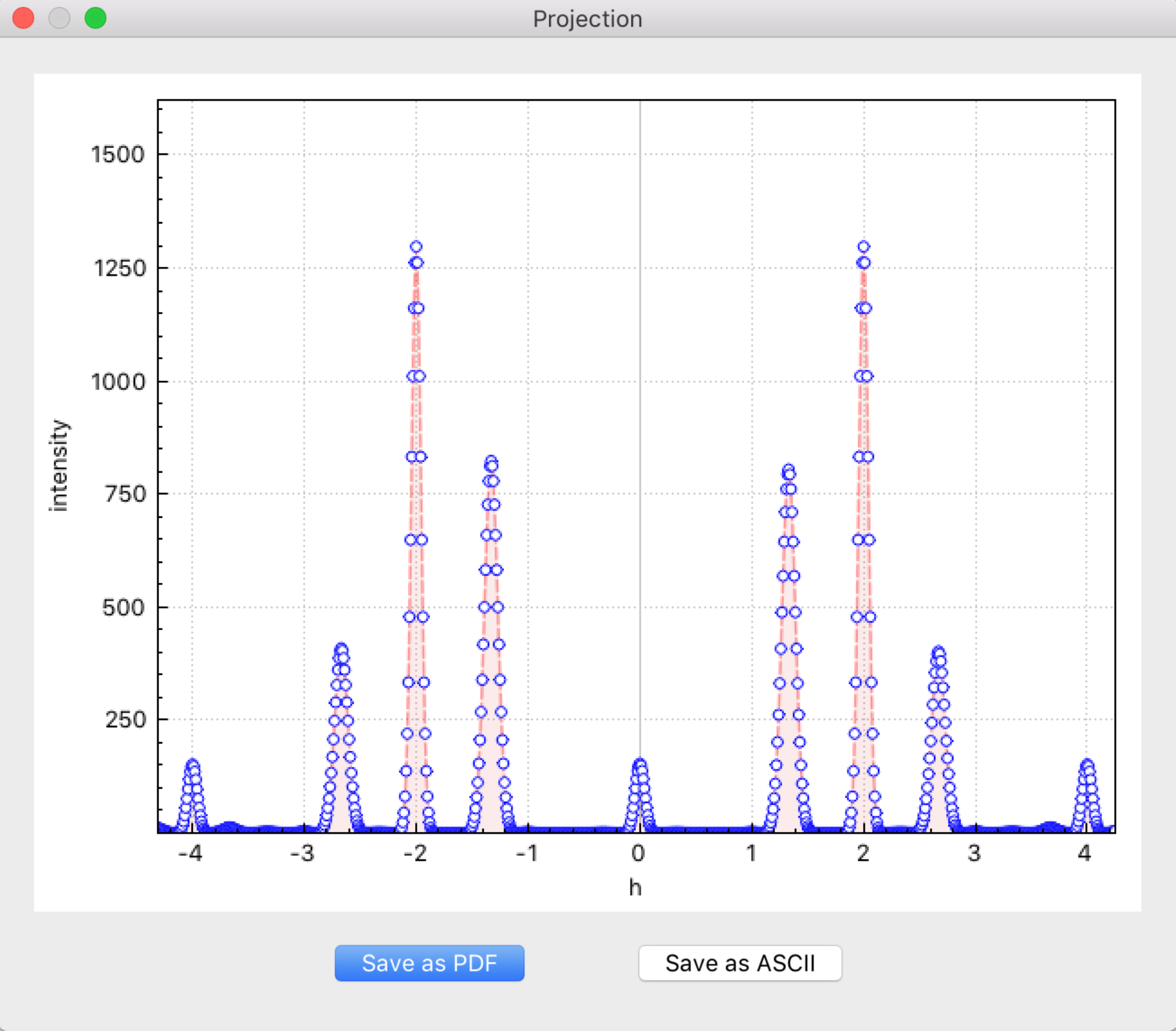}
	\label{fig:maps} 
\end{figure}

Besides the correct description of a magnetic structure, \textsc{Mag2Pol} needs information concerning the experimental geometry in order to calculate structure factors and polarization matrices. Therefore, the wavelength should be given in \AA ngstr\"oms, which already allows the calculation of structure factors (note that although structure factors are independent from the wavelength, \textsc{Mag2Pol} considers it to distinguish from inaccessible reflections). The calculation takes into account the scale and extinction parameters as well as the different magnetic domains and structural twins with their corresponding populations which are adjustable in the respective \textit{Structure} menu entries. The calculation of the polarization matrix depends on the orientation of the sample, which can be given either by defining the vertical sample axis (as a reciprocal lattice vector) or by supplying the orientation matrix in the case of a Cryocradle experiment. For the latter, the orientation angles $\chi$ and $\varphi$ are automatically calculated using the formalism of Ref.~\onlinecite{bus1967} and shown in the respective boxes. The program furthermore offers the possibility to calculate an orientation (UB) matrix from the centering angles of observed reflections in Cryocradle geometry with an optional refinement of the lattice parameters respecting the crystal symmetry.\footnote{The conventions of the D3 (ILL, Grenoble) \cite{d3} instrument are employed, where the $y$ axis is parallel to the beam, $z$ is vertical pointing downwards and $x$ completes the right-handed coordination system. The angles $\gamma$, $\omega$ and $\varphi$ (the latter for $\chi=0$) are counted negative for clockwise rotations when viewing from the top, $\chi$ is negative for clockwise rotations when viewing towards the source (for $\omega$=$\varphi$=0)} \newline
The calculation of structure factors and polarization matrices is done for the Bragg reflection ($hkl$)$\pm\mathbf{q}$ entered by the user. If the propagation vector is 0, there is no difference concerning the calculation of a $\pm\mathbf{q}$ satellite. If the checkbox \textit{show} is activated, the local coordination system (see Sec.~\ref{sec:matrices}) and the magnetic interaction vector $\mathbf{M}_\perp$ are drawn.\newline
The results of the calculation are shown in the output text window and the individual polarization matrix entries are shown in their respective boxes in the \textit{Geometry} section of the window.\newline
Whole lists of reflections can be generated for given criteria like $hkl$ range, intensity, polarization and diffractometer angles. This can be useful as a preparation for an experiment e.g. by selecting those magnetic reflections which show a large chiral component in the $P_{yx}$ term. The reflections which fullfil the given criteria are shown in a table which can be ordered according to a specific parameter. Intensity maps can be generated for which all nuclear and magnetic structure factors for integer and satellite reflections will be calculated in the range from -10 to 10 for $h$, $k$ and $l$. The resolution corresponding to the full width at half maximum of a Gaussian function can be set individually for the $h$, $k$ and $l$ directions. A logarithmic scale can be chosen or nuclear scattering can be hidden. Plot controls like zooming, color scale and data cursor are enabled. Using a key modifier while zooming will automatically calculate a one-dimensional projection along a reciprocal space direction corresponding to the drawn rectangle (see. Fig.\ref{fig:maps}). \newline In case of multi-q magnetic structures scattering from differently modulated components of the magnetic structures can be added. For that the magnetic structure in the main window should be modified (especially the propagation vector) and regenerated. Afterwards, the scattering from that new magnetic structure can be added to the intensity map. \newline Intensity maps can be saved as pdf or ASCII files by clicking the corresponding button at the bottom.

\subsection{Refinement}

\begin{figure}
	\centering
	\includegraphics[width=0.48\textwidth]{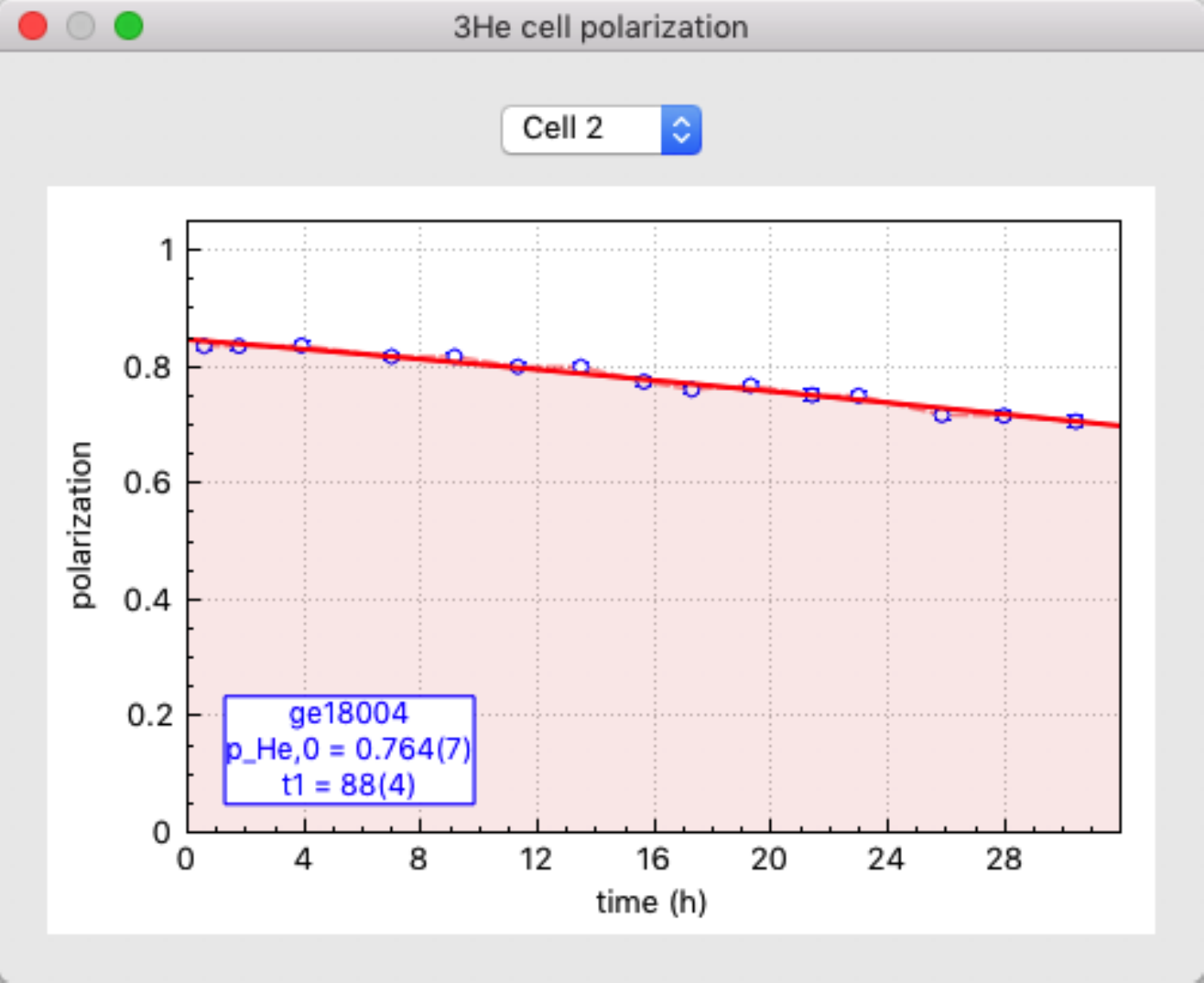}
	\caption{Visualization of the spin-filter efficiency decay as a function of time.}
	\label{fig:he3cell} 
\end{figure}

The main purpose of \textsc{Mag2Pol} is the refinement of a magnetic structure model to both SNP data and/or to integrated intensities from a monochromatic single-crystal diffraction experiment. After the structural data, the magnetic structure model, magnetic domains, scale factor and extinction parameters have been entered and the unit cell has been successfully generated and rendered, one can proceed to the refinement module. Note that in the present version, the loading routines for polarized neutron data are only applicable for D3 data file formats, however, other polarized neutron diffractometers, e.g. POLI (FRM-II, Garching) \cite{poli}, 5C1 or 6T2 (LLB, Saclay) \cite{5C1,6T2} will be included in the near future. Polarization matrices can be loaded in the D3 \texttt{*.fli} format containing $(hkl)$ indices, polarization channels and values as well as information about the used $^3$He spin filters (the files used for the analysis of the example shown in Sec.~\ref{sec:ex1} can be obtained from the \textsc{Mag2Pol} download site, see Sec.~\ref{sec:availability}).  \newline
At this point the $^3$He spin-filter efficiency should be taken into account. Usually the efficiency is monitored during a SNP experiment by measuring e.g. the polarization $P_{zz}$ on a purely nuclear Bragg peak, which is known to be 1, several times a day. The reduction from the expected value is due to the imperfect initial polarization of the neutron beam $P_0$ and the spin-filter efficiency. In the case of a loaded \texttt{*.fli} file the program analyzes the lines which signal a cell change and the time of the respective observations to show the decrease of efficiency as a function of hours. On D3 - depending on the used cell - the decay of neutron polarization is usually between 10 and 20\% per day. The reflection on which the calibration has been performed throughout the experiment and the polarization channel has to be selected before starting the calculation. The polarization data $P_{mes}(t)$ measured at the time $t$ will then be corrected with respect to the decaying spin-filter efficiency $P_{He}(t)$ and the polarization of the incoming neutron beam $P_0$ according to:
\begin{equation}
P_{corr} = P_{mes}(t)\cdot \frac{P_0}{P_{He}(t)}
\end{equation}
and repeated reflections can optionally be merged. A more precise merging is possible, if the respective raw data files corresponding to the reflection index numbers (Numor) are put in the same folder as the \texttt{*.fli} file. Like this the individual counts (background - peak - background) can be summed up and weighted by the respective counting times. Otherwise, a weighted average will be calculated from the standard deviations given in the \texttt{*.fli} file. In order to interpolate the spin-filter efficiency for any given time an exponential decay function of the form: 
\begin{equation}
P_{He}(t) = \tanh[O\cdot P_{He}(t=0) \cdot \exp(-t/T_{1})]
\end{equation}
is used, where $O=\lambda\cdot l \cdot p\cdot 7.282\cdot 10^{-2}$ is the cell opacity at room temperature (see e.g. Ref.~\onlinecite{sur1997}). The wavelength and the cell pressure $p$ are fixed, while the halflife $T_{1}$ and the initial polarization of the cell $P_{He}(t=0)$ are refined. The spin-filter efficiency and the refined values can be visualized like shown in Fig.~\ref{fig:he3cell}.
The different cells recognized from the entries in the \texttt{*.fli} file can be chosen from the combo box. The legend shows the name of the cell as well as the refined values including standard deviations for the initial cell polarization and the halflife.\newline 
The corrected data can be saved from the fit window in order not to repeat this procedure.\newline 
Note that when chiral or nuclear-magnetic interference scattering is present, the polarization data should not be corrected as the respective entries contain terms which are dependent on the initial neutron polarization and terms which are not (see Blume-Maleev equations). In that case the checkbox \textit{Refine on uncorrected data using cell efficiency} should be activated before applying the correction. The program will still take into account the cell efficiency at a given time of an observation, but it will be used in combination with the initial neutron polarization to calculate the respective polarization and analysis axes as described in Sec.~\ref{sec:matrices}. In this case no merging of reflections is possible.\newline

Loading D3 raw data Numors directly is intended for a special type of measurement, where the initial neutron spin is rotated within the local $y$-$z$ plane. The positions and amplitudes of the minima and maxima allow important conclusions concerning the magnetic structure. In this case \textsc{Mag2Pol} will calculate the polarization directions according to the CRYOPAD currents which were set to rotate the polarization and which are stored in the Numors. Furthermore, the polarization value including the time of the observation extracted from the Numor files. As no information is present concerning any potential spin-filter cell changes only those data should be loaded which have been measured with the same cell. The correction with respect to the cell efficiency is done as described above (note that for a final polarization vector within the $y$-$z$ plane no chiral scattering is present).  \newline

Integrated intensity data can optionally be added in order to perform a joint refinement together with SNP data, which combines the sensitivity to the spin amplitude of the first data set with the high accuracy concerning the spin alignment of the latter data set. It is also possible to refine the nuclear structure (including scale and extinction factors) either on its own or together with the magnetic structure, but - for the moment - with integrated intensity data only. The supported data types correspond to \textsc{FullProf} \texttt{*.int}-files containing the $(hkl)$ reflections, integrated intensities and standard deviations either without propagation vector information ($\mathbf{q}=0$) or by stating two propagation vectors to be added to or subtracted from the integer $(hkl)$ reflections. The \texttt{*.col} output files of the \textsc{RACER} integration program \cite{racer} and the \texttt{*.fsq} output files of \textsc{COLL5} \cite{leh1974} can directly be opened with \textsc{Mag2Pol} which allows the user to merge the data with respect to a given space group and apply an absorption correction based on a spherical or cylindrical sample shape following the anlytical method reported by~\onlinecite{hu2012}. Data statistics and the merged data set will then be given. Data sets in \texttt{*.col} and \texttt{*.fsq} format can be subtracted from each other, e.g. a low-temperature $\mathbf{q}=0$ data set and a high-temperature nuclear background. If purely magnetic scattering data is treated the corresponding checkbox can be activated, for which the program will ignore the nuclear structure factor and the given structure model can be kept unchanged.  \newline
Flipping ratios are also loaded in the D3 \texttt{*.fli} format, where the necessary information are the ($hkl$) indices, the flipping ratios and their error. The example file used for the analysis in Sec.~\ref{sec:ex3} can be obtained from the \textsc{Mag2Pol} download site, see Sec.~\ref{sec:availability}.\newline
The parameters to be refined are chosen by simply ticking checkboxes next to the parameters' values and linear constraints of the form $p_1 = a\cdot p_2+b$ can be added. The refinement algorithm used is the non-linear optimization module within the \textsc{Eigen} library, which itself is a port of \textsc{Minpack}, a robust and well renowned Fortran package. In the case of a correlated refinement between polarization matrices and integrated intensities a weighting factor $w$ can be adjusted, for which the least-squares refinement will minimize:
\begin{equation}
\chi_w^2 = w\cdot\chi^2_{Int} + (1-w)\cdot\chi^2_{Mat},
\end{equation}
where $\chi^2_{Int}$ and $\chi^2_{Mat}$ are the agreement factors of the integrated intensities and the polarization data, respectively. Besides the maximum number of iterations the convergence criterium can be set in the program's settings and it refers to the maximum relative shift of a parameter in an iteration (note that all values in the \textsc{Mag2Pol} settings menu are saved in the program registry). For every iteration the $\chi^2$ value will be printed together with the maximum relative parameter shift. Either when the maximum number of iterations has been reached or the convergence criterium is fullfilled, the refinement will stop and list the $\chi^2$, reduced $\chi^2$ values (and $R_F$ for integrated intensities) together with the refined parameters and their respective standard deviations. The latter are obtained from the diagonal of the raw-correlation matrix:
\begin{equation}
\mathbf{C} = \boldsymbol{\alpha}^{-1}\cdot \chi^2_r
\end{equation}
$\boldsymbol{\alpha}$ is a $n\times n$ square matrix with $n$ the number of refined parameters $x_j$:
\begin{equation}
\boldsymbol{\alpha} = \mathbf{D}^T\cdot\mathbf{D}
\end{equation}
with $\mathbf{D}$ being a $m\times n$ matrix ($m$ is the number of observations $y_i$ or calculated values $y_{c,i}$) where the $j^{th}$ column is given by:
\begin{equation}
\mathbf{D}^{(j)}= \frac{y_{c,i}(x_j+\Delta)-y_{c,i}(x_j-\Delta)}{2\Delta \sigma_i}
\end{equation}
The reduced $\chi^2$ value is defined as:
\begin{equation}
\chi^2_r = \frac{\chi^2}{(m - n)}.
\end{equation}

All observed and calculated data points are shown graphically with usual zoom controls and data cursor support and can be exported into a pdf or an \text{ASCII} file. By accepting the refinement the new parameters will be transferred to the main window and the visualization of the magnetic structure will be re-rendered. By simply closing the fit window without accepting, the initial parameters will be restored. It is also possible to use an undo function which permits the user to go back step by step and reload all parameters and refinement flags at each stage before the \textit{Fit} button has been clicked (useful when the refinement diverges or becomes considerably worse).

\section{Theory}
\label{sec:theory}

\subsection{Conventions and structure factors}
\label{sec:conv}

\textsc{Mag2Pol} uses the plane wave function following conventions of quantum mechanics, which is generally used for neutron scattering and which is complex conjugate of that used in X-ray scattering:
\begin{equation}
\psi_{QM}=\psi_{XR}^*=e^{i(\mathbf{k}\mathbf{r}-\omega t)}
\end{equation}
which requires:
\begin{equation}
b=b'-ib'' \quad with \quad b''>0
\end{equation}
for the scattering length and yields the scattered wave as a superposition of the plane wave and the spherical wave in the form of:
\begin{equation}
\psi(\mathbf{r})=e^{i\mathbf{k}\mathbf{r}}+f(\theta,\phi)\frac{e^{ikr}}{r}.
\end{equation}
The scattering amplitude is the Fourier transform of the scattering potential and is given by:
\begin{equation}
f(\theta,\phi) \propto - \int V(\mathbf{r}) e^{-i\mathbf{Q}\mathbf{r}}d\mathbf{r}
\end{equation}
with the scattering vector $\mathbf{Q}=\mathbf{k}_f-\mathbf{k}_i$.
Note that using the conventions for X-ray scattering or $\mathbf{Q}=\mathbf{k}_i-\mathbf{k}_f$ leads to the complex conjugate of the scattering amplitude. 
The nuclear structure factor can therefore be written as:
\begin{equation}
\label{eq:fn}
N(\mathbf{Q}) = \sum\limits_j o_j b_j e^{-2\pi i (hx_j+ky_j+lz_j)}e^{-B_j\sin^2\theta/\lambda^2}
\end{equation}
with the second exponential term being the Debye Waller factor ($B_j$ is the isotropic temperature factor). In the case of anistropic temperature factors the Debye Waller factor is given by:
\begin{equation}
DW = e^{-\mathbf{Q}\boldsymbol{\beta}\mathbf{Q}}
\end{equation}
where $\boldsymbol{\beta}$ is the anisotropic temperature tensor.
The sum in Eq.~\ref{eq:fn} is done over all the atoms in the unit cell and $o_j$ denotes the occupation of atom $j$. Note that the leading minus sign has been dropped as the square of the structure factor (or the nuclear-magnetic interference) is used in all calculations. \newline The magnetic structure factor writes as:
\begin{equation}
\mathbf{M}(\mathbf{Q}) = \sum\limits_j p_n o_j \mathbf{S}_j f_j(\mathbf{Q}) e^{-2\pi i (hx_j+ky_j+lz_j)}e^{-B_j\sin^2\theta/\lambda^2}
\end{equation}
where $p_n=\frac{\gamma_n r_e}{2}$ is the conversion factor between Bohr magnetons and scattering lengths with $\gamma_n$ being the neutron magnetic moment in nuclear magnetons and $r_e$ the classical electron radius. A factor 0.5 is applied, when the non-zero propagation vector is in the interior of the first Brillouin zone. $\mathbf{S}_j$ are the Fourier coefficients of the magnetic moment expansion according to: 
\begin{equation}
\boldsymbol{\mu} = \frac{1}{2} \sum\limits_\mathbf{q} S_\mathbf{q}e^{i\mathbf{q}\mathbf{R}}+S^*_\mathbf{q}e^{-i\mathbf{q}\mathbf{R}}
\end{equation}
The Fourier coefficients of a given magnetic moment contain a phase factor determined by symmetry and a refinable one.\newline
The calculated intensities, which are to be compared to the observed ones, are given by: 
\begin{equation}
I_N(\mathbf{Q}) = scale\cdot y\cdot NN^*
\end{equation}
and:
\begin{equation}
I_M(\mathbf{Q}) = scale\cdot y \cdot \mathbf{M_\perp}(\mathbf{Q})\mathbf{M_\perp}^*(\mathbf{Q})
\end{equation}
with the magnetic interaction vector:
\begin{equation}
\mathbf{M}_\perp = \mathbf{\hat{Q}}\times(\mathbf{M}(\mathbf{Q})\times \mathbf{\hat{Q}}).
\end{equation}
$scale$ is a simple scaling factor and $y$ is the extinction correction: 
\begin{equation}
y=\left(1+\frac{0.001x_{aniso}F_c^2\lambda^3}{4\sin(2\theta)(\sin\theta/\lambda)^2}\right)^{-\frac{1}{2}}
\end{equation}
which is also implemented in \textsc{FullProf} \cite{fullprof} and \textsc{ShellX} \cite{shelx}, and which is most similar to the treatment shown in \cite{lar1970}. Herein, $x_{aniso}$ is the anisotropic extinction parameter and $F_{c}^2$ is the calculated squared structure factor. $x_{aniso}$ is obtained by operating the tensor $\mathbf{x}$ subsequently on the scattering vector ($hkl$):
\begin{equation}
\begin{split}
x_{aniso}&=\left[\begin{pmatrix} x_{11} && x_{12} && x_{13} \\ 0 && x_{22} && x_{23} \\ 0 && 0 && x_{33} \end{pmatrix}\begin{pmatrix}h \\ k \\ l \end{pmatrix}\right]\begin{pmatrix}h \\ k \\ l \end{pmatrix}\\ &=x_{11}h^2+x_{22}k^2+x_{33}l^2+x_{12}hk+x_{13}hl+x_{23}kl
\end{split}
\end{equation}
\newline\newline 

\subsection{Polarization matrices}
\label{sec:matrices}

In addition to the general law in magnetic neutron scattering that only the perpendicular component of the magnetic structure factor (the magnetic interaction vector $\mathbf{M}_\perp$) contributes to magnetic scattering, neutrons with their polarization axis parallel to $\mathbf{M}_\perp$ will undergo a non-spin-flip (NSF) scattering process, while neutrons with polarization perpendicular to $\mathbf{M}_\perp$ will be scattered with a spin-flip (SF). Chiral scattering is polarization dependent and will create additional polarization along the scattering vector $\mathbf{Q}$ and can therefore be revealed by analyzing the final neutron polarization along that direction. Based upon those laws a standard local coordination system is employed in polarized neutron scattering with polarization analysis, where $\mathbf{x}$ is parallel to the scattering vector $\mathbf{Q}$, $\mathbf{z}$ is the vertical direction of the diffractometer and $\mathbf{y}$ completes the right-handed coordination system. \newline 
As mentioned in Sec.\ref{sec:Introduction} the scattered intensity and final polarization of the neutrons can be calculated using the Blume-Maleev equations. A much more elegant way (and far more suitable for matrix-based calculations) is based on the density matrix formalism \cite{sch2002}, where the expectation value of the mixed state $P_{ij}=\frac{n^+-n^-}{n^++n^-}$ ($n^+$ and $n^-$ are the numbers of detected spin-up and spin-down neutrons) is obtained by:

\begin{equation}
P_{ij} = \frac{\Tr(\boldsymbol{\sigma}\cdot \mathcal{M}\cdot \boldsymbol{\rho}\cdot \mathcal{M}^\dagger)}{\Tr(\mathcal{M}\cdot \boldsymbol{\rho}\cdot \mathcal{M}^\dagger)}
\end{equation}
with the density matrix representing the final beam:
\begin{equation}
\boldsymbol{\sigma} = \begin{pmatrix}
P_{f,z} & P_{f,x}-iP_{f,y} \\
P_{f,x}+iP_{f,y} & -P_{f,z}
\end{pmatrix},
\end{equation}
and the density matrix representing the initial beam:
\begin{equation}
\boldsymbol{\rho} =\frac{1}{2} \begin{pmatrix}
1+P_{i,z} & P_{i,x}-iP_{i,y} \\
P_{i,x}+iP_{i,y} & 1-P_{i,z}
\end{pmatrix}.
\end{equation}
$\mathcal{M}$ represents the scattering system:
\begin{equation}
\mathcal{M} = \begin{pmatrix}
M_{\perp,z} & -iM_{\perp,y} \\
iM_{\perp,y} & -M_{\perp,z}
\end{pmatrix}
\end{equation}
and $\mathcal{M}^\dagger$ denotes the conjugate transpose of $\mathcal{M}$. $\mathbf{P}_i$ and $\mathbf{P}_f$ are the polarization and analysis axes of the incident and final neutrons, whose modules are smaller or equal to 1. In fact, the imperfect neutron polarization $P_0$ is contained in $\mathbf{P}_i$, while the spin-filter efficiency of a $^3$He cell is multiplied to the unit vector along $\mathbf{P}_f$. Those polarization and analysis imperfections therefore define the matrices $\boldmath{\sigma}$ and $\boldmath{\rho}$. By aligning the vectors $\mathbf{P}_i$ and $\mathbf{P}_f$ along $x$, $y$ and $z$, respectively, a total of nine observations can be made which define the polarization tensor as a combination of the rotational part $\mathcal{P}$ and the created/annihilated polarization $\mathbf{P}'$. The final neutron polarization $\mathbf{P}_f$ is therefore related to the incident polarization $\mathbf{P}_i$ by:
\begin{equation}
\mathbf{P}_f = \mathcal{P}\mathbf{P}_i+\mathbf{P}'
\end{equation} 
However, the expectation values obtained by the density matrix approach are not limited to the cases where the incident and final polarization axes are along the local $x$, $y$ and $z$ directions. In fact any direction can be used and this is exploited by \textsc{Mag2Pol} to calculate the polarization values for rotations of $\mathbf{P}_i$ within the $y$-$z$ plane. 

\subsection{Flipping ratios}
\label{sec:frs}

The flipping ratio $R$ of a Bragg reflection is defined as the ratio between the diffracted intensities for incident spin-up and spin-down neutrons and is given by: 
\begin{equation}
R = \frac{I^+}{I^-}=\frac{NN^* + (N\mathbf{P}_0\mathbf{M}_\perp^* + N^*\mathbf{P}_0\mathbf{M}_\perp) + M_\perp M_\perp^*}{NN^* - (N\mathbf{P}_0\mathbf{M}_\perp^* + N^*\mathbf{P}_0\mathbf{M}_\perp) + M_\perp M_\perp^*}
\end{equation}
where $\mathbf{P}_0$ is the vector of the initial neutron polarization ($P_0\leq 1$) to be taken along the vertical axis of the diffractometer and the dot product $\mathbf{P}_0\mathbf{M}_\perp$ gives the projection of the magnetic interaction vector onto the vertical polarization axis \cite{bro2004}.  \newline \newline 
The extinction correction for flipping ratios corresponds to the one which is applied in \textsc{FullProf} \cite{fullprof}, see the \textit{Flipping Ratios in FullProf} manual. The flipping ratio is then written as: 
\begin{equation}
R=\frac{Ip^+_p+(N^*\mathbf{p}^+_m\mathbf{M}_\perp + N\mathbf{p}^+_m\mathbf{M}_\perp^*)+(1-q^2)M_\perp M_\perp^*y_{pm}}{Ip^-_p+(N^*\mathbf{p}^-_m\mathbf{M}_\perp + N\mathbf{p}^-_m\mathbf{M}_\perp^*)+(1-q^2)M_\perp M_\perp^*y_{pm}} 
\end{equation}
with $I=NN^*+M_{\perp,z} M_{\perp,z}^*$ and the correction factors:
\begin{equation}
p^\pm_p=\frac{1}{2}\left[(1\pm P_0)y_p+(1\mp
P_0)y_m\right]
\end{equation}
\newline
\begin{equation}
\mathbf{p}^\pm_m=\frac{1}{2}\left[(1\pm P_0)y_p-(1\mp
P_0)y_m\right]\mathbf{\hat{P}}_0
\end{equation}
\newline
\begin{equation}
y_{p}=(1+\frac{0.001 x_{aniso}I^+ \lambda^3}{4(\sin\theta/\lambda)^2\sin(2\theta)})^{-1/2},
\end{equation}
\newline
\begin{equation}
y_{m}=(1+\frac{0.001 x_{aniso}I^- \lambda^3}{4(\sin\theta/\lambda)^2\sin(2\theta)})^{-1/2},
\end{equation}
\newline
\begin{equation}
y_{pm}=(1+\frac{0.001 x_{aniso}(1-q^2)M_\perp M_\perp^* \lambda^3}{4(\sin\theta/\lambda)^2\sin(2\theta)})^{-1/2},
\end{equation}
\newline

where $P_0$ is the beam polarization ($\mathbf{\hat{P}}_0$ is a unit vector along the neutron polarization axis) and $I^\pm$ are the uncorrected
intensities for a spin-up and spin-down beam, respectively. $q=\sin\alpha$ with $\alpha$ being the angle between the scattering vector and the vertical axis.

\subsection{Multipole expansion}
\label{sec:mpoles}

The magnetization density $m(\mathbf{r})$ is described by: 
\begin{equation}
m(\mathbf{r}) = R'_0 C'_0{^0} d'_0{^0}(\mathbf{\hat{r}}) + \sum\limits_{l=0}^4 R_l(r) \sum\limits_{m=-l}^lC_l^m d_l^m(\mathbf{\hat{r}})
\end{equation}
where the $d_l^m$ are the density-normalized real spherical harmonics whose normalization constants can be found in \onlinecite{han1978}. The primed quantities denote the first monopole besides the one specified by $R_0$, $C_0^0$ and $d_0^0$. $R_l$ is the radial dependence of a Slater-type orbital: 
\begin{equation}
R_l(r) = \frac{Z^{n_l+3}}{(n_l+2)!}r^{n_l}\exp(-Z_lr)
\end{equation}
where $n_l$ and $Z_l$ can be chosen to be different for each $l$ value. $Z_l$ and the population coefficients $C_l^m$ are refinable parameters.
The normalization is such that: 
\begin{equation}
\int\limits_{r=0}^\infty R_lr^2 dr= 1.
\end{equation}
The wave-function normalized real spherical harmonics $y_l^m$ are obtained from the spherical harmonics $Y_l^m$ according to: 
\begin{equation}
y_l^m = 
\begin{cases}
\sqrt{2}\operatorname{Im}( Y_l^{\left|m\right|} ),& \text{if } m < 0\\
Y_l^0,              & \text{if } m = 0 \\
\sqrt{2}\operatorname{Re}( Y_l^{\left|m\right|} ),& \text{if } m > 0
\end{cases}
\end{equation}
The spherical harmonics parametrized in the angles $\theta$ and $\varphi$ are given by:
\begin{equation}
Y_l^m(\theta,\varphi) = \sqrt{\frac{2l+1}{4\pi}\cdot\frac{(l-m)!}{(l+m)!}}P_l^m(\cos\theta)e^{im\varphi}
\end{equation}
As indicated e.g. in \onlinecite{mic2015} the Condon-Shortley phase $(-1)^m$ which can be seen in some textbooks is included in the associated Legendre polynomials $P_l^m$ so that, e.g. $P_1^{1}(x)=-(1-x^2)^\frac{1}{2}$ and $P_1^{-1}(x) = -\frac{1}{2}P_1^1(x)$. Therefore, the density-normalized real spherical harmonics are normalized so that:
\begin{equation}
\int\limits_{\theta=0}^\pi\int\limits_{\varphi = 0}^{2\pi} \left|d_l^m(\theta,\varphi)\right|\sin(\theta) d\varphi d\theta= 
\begin{cases}
1, & \text{for } l=0 \\
2, & \text{for } l > 0
\end{cases}
\end{equation}
The magnetic form factor is given by the Fourier transform of the magnetization density and results in: 
\begin{equation}
f(\mathbf{Q}) = \langle j'_l(Q)\rangle C'_0{^0} d'_0{^0}(\mathbf{\hat{Q}}) + \sum\limits_{l=0}^4 i^l \langle j_l(Q)\rangle \sum\limits_{m=-l}^l C_l^m d_l^m(\mathbf{\hat{Q}})\ 
\end{equation}
where $\mathbf{\hat{Q}}$ is parametrized in terms of $\theta$ and $\varphi$ and $\langle j_l(Q) \rangle$ being the Fourier-Bessel transform of the radial density:
\begin{equation}
\langle j_l(Q) \rangle = \int\limits_{r=0}^\infty R_l(r)j_l(Qr)4\pi r^2dr
\end{equation}
$j_l(x)$ are the spherical Bessel functions. Note that $f(\mathbf{Q}$=$0)$=$1$ for $C_0^0=1$, since $\langle j_0(0) \rangle=4\pi$ and $d_0^0(\mathbf{\hat{Q}})=\frac{1}{4\pi}$.

\section{Examples}
\label{sec:results}

\subsection{Complex magnetic ordering in Swedenborgites}
\label{sec:ex1}

The Swedenborgite family represents a compound class with a large variety of physical properties ranging from long-range magnetic order \cite{rei2014} to spin-glass \cite{val2002,val2004} and spin-liquid behaviour \cite{val2006,sch2007,val2011}, which is a result of the geometrical frustration in this layered kagome system and the type of distortion relieving it. CaBaCo$_3$FeO$_4$ is observed to have an orthorhombic crystal structure and orders antiferromagnetically below $T_N\approx 140$ K \cite{val2009b}. The magnetic structure was investigated by measuring integrated intensities by standard neutron diffraction and polarization matrices by spherical neutron polarimetry experiments. Usual spin configurations on a kagome lattice could not explain the integrated intensity data, however, close inspection of the polarization matrices suggested a symmetry reduction to the magnetic superspace symmetry $P2_1'$. Both data sets could be simultaneously analyzed using the \textsc{Mag2Pol} program yielding a complex magnetic structure which fits the experimental data excellently \cite{qur2018b}. Hereby, the program's ability to easily generate magnetic form factors for magnetic sites which are shared by different atoms was exploited. Furthermore, the number of refined parameters could be substantially reduced by setting linear constraints due to the magnetic symmetry. The resulting spin configuration can in fact be mapped onto the classical $\sqrt{3}\times\sqrt{3}$ structure of a kagome lattice with the difference that the spins do not rotate within the $a$-$b$ plane, but around an axis close to the [110] direction. The population of the three structural twins and four magnetic domains (two orientational and two chirality domains) was found to be homogeneous as one would expect without inducing any preference by external parameters. \newline 
The files used for the analysis of this example can be obtained from the \textsc{Mag2Pol} download site, see Sec.~\ref{sec:availability}.

\subsection{Cycloidal magnetic ordering and magnetic domain population in the hybrid multiferroic (NH$_4$)$_2$[FeCl$_5$(D$_2$O)]}
\label{sec:ex2}
The family of erithrosiderite-type compounds A$_2$[FeX$_5$(H$_2$O)] \cite{car1985}  where A stands for an alkali metal or ammonium ion and X for a halide ion, represents a new route to obtain materials with strong magneto-electric coupling. Some of the compounds of this family are linear magneto-electric materials, while (NH$_4$)$_2$[FeCl$_5$(H$_2$O)] \cite{ack2013} is an actual spin-driven multiferroic in its ground state. The mechanisms of multiferroicity in this compound have been determined, in different zones of its magnetic field-temperature phase diagram, on the basis of its crystal and magnetic structures obtained by neutron diffraction \cite{rod2015,rod2017}. The magnetic structure in the ground state (below 6.9 K and in absence of external magnetic field) is cycloidal, propagating along the $c$ axis and with magnetic moments mainly contained in the $a$-$c$ plane. The ferroelectric polarization, primarily directed along the $a$ axis, observed in this phase, is basically explained through the spin current mechanism \cite{rod2015}. Neutron spherical polarimetry has been used to determine the absolute magnetic configuration and domain population of this system under different external electric fields. The cycloidal magnetic structure of the ground state implies two chiral domains with opposite rotation of the cycloids. By applying an electric field, the population of the chiral magnetic domains can be manipulated and therefore the electric polarization, and it was shown that the system is completely switchable, which represents a direct evidence of its multiferroic character. \textsc{Mag2Pol} was used to analyze the domain populations from polarization matrices recorded in different electric-field cooling conditions. Spherical neutron polarimetry data were also combined in a joint fit with integrated intensities from standard unpolarized neutron diffraction to refine the previously proposed magnetic structure of the ground state, obtaining a more accurate magnetic structure model, allowing the refinement of the ellipcity and inclination of the rotation plane of the cycloids \cite{rod2018}. 

\subsection{Spin density of an end-to-end azido double-bridged Cu$^{II}$ dinuclear complex }
\label{sec:ex3}
The end-to-end azido double-bridged copper(II) complex [Cu$_2$L$_2$(N$_3$)$_2$] (L = 1,1,1-trifluoro-7-(dimethylamino)-4-methyl-5-aza-3-hepten-2-onato) was investigated by density functional theory calculations and polarized neutron scattering revealing that the spin density is mainly localized on the copper(II) ions with a small degree of delocalization on the ligand (L) and terminal azido nitrogens \cite{aro2007}. The analysis in terms of a Cu 3$d$ orbital model yielded an important contribution of the Cu $d_{x^2-y^2}$ orbital and a small population of the $d_{z^2}$ orbital.  \textsc{Mag2Pol} was tested on the same data set and the obtained results were consistent concerning the overall magnetic moment on the involved ions (Cu, Ni, O) and the multipole occupation coefficients of the Cu ion. As previously reported, an important population on the Cu $d_{x^2-y^2}$ and $d_{z^2}$ orbitals is observed which corresponds to significant coefficients for the monopole ($C_0^0$), quadrupole ($C_2^0$, $C_2^2$) and hexadecapole ($C_4^0$, $C_4^4)$ populations according to the generalized relations between $d$-orbital occupations of transition-metal atoms and multipole population parameters \cite{hol1983}. The small deviations within one $\sigma$ interval stem from the fact that the first monopole of Cu had to be approximated by a Slater-type orbital ($n$=1, $Z$=6.31 a.u.) instead of using a table of values corresponding to ($\langle j_0\rangle$+$\langle j_2\rangle$) which was used to correct for the orbital contribution in the original paper. The example files can be obtained at the \textsc{Mag2Pol} download site, see Sec.~\ref{sec:availability}.

\section{Computational requirements, availability and documentation}
\label{sec:availability}

The \textsc{Mag2Pol} source code is entirely written in C++ and is mainly based on the \textsc{Qt5}, \textsc{Eigen}, \textsc{QCustomPlot} and \textsc{OpenGL} libraries. It is deployed as a compiled application (under the GNU Lesser General Public License version 3) for the main platforms Windows, MacOSX and Linux without any dependency as all necessary libraries are either included in the package (MacOSX/Windows) or are downloaded upon installation (using gdebi under Linux) and dynamically linked to the progam. In order to run under Windows, \textsc{Qt5} depends on Microsoft Visual Studio, for which the Visual Studio Redistributable Packages are included and need to be installed prior to the use of \textsc{Mag2Pol}. \newline The program can be downloaded from the ILL D3 instrument webpage under \textit{Software}: https://www.ill.eu/instruments-support/instruments-groups/instruments/d3/software. A short description of the program, installation guides and contact information are mentioned as well. The software comes with an extensive manual which can be accessed from within the application. 

\section{Conclusion}
\label{sec:conclusion}

\textsc{Mag2Pol} is a unique GUI-based program to refine magnetic structure models and magnetic domain populations against spherical neutron polarimetry data with the possibility to employ a correlated refinement by including an additional integrated intensity data set. The treatment of integrated intensities alone as well as the treatment of flipping ratios with optional multipole expansion of the magnetic form factor is supported. The correct calculation of nuclear and magnetic structure factors as well as flipping ratios was verified for different data sets against the \textsc{FullProf} code and the application of \textsc{Mag2Pol} was shown to be successful in different examples. The intuitive interaction with the program and its graphical output - e.g. magnetic structures in different domains, orientation of local reference frame, magnetic interaction vector, multipoles - is intended to facilitate the approach of even unexperienced users to the rather complex technique of spherical neutron polarimetry. Features like the generation of reflection lists, intensity maps and cuts along $Q$ directions may serve as a useful preparation of neutron scattering experiments.

\section{Acknowledgements}
\label{sec:ack}

Invaluable discussions with L. C. Chapon, J. Rodr\'iguez-Carvajal and B. Gillon are greatly appreciated. 
\appendix

\end{document}